# A rare simultaneous detection of a mid-latitude plasma depleted structure in O($^1$D) 630.0 nm and O($^1$S) 557.7 nm all-sky airglow images on a geomagnetically quiet night

D. Patgiri[1], R. Rathi[1], V. Yadav[2], D. Chakrabarty[3], M. V. Sunil Krishna[1,4], S. Kannaujiya[5], P. Pavan Chaitanya[6], A. K. Patra[6], Jann-Yenq Liu[7,8,9], S. Sarkhel[1,4,*]

[*]Sumanta Sarkhel, Department of Physics, Indian Institute of Technology Roorkee, Roorkee - 247667, Uttarakhand, India (sarkhel@ph.iitr.ac.in)

[1]Department of Physics, Indian Institute of Technology Roorkee, Roorkee – 247667, Uttarakhand, India

[2]Aryabhatta Research Institute of Observational Sciences, Nainital – 263001, Uttarakhand, India

[3]Space and Atmospheric Sciences Division, Physical Research Laboratory, Ahmedabad, 380009, Gujarat, India

[4]Centre for Space Science and Technology, Indian Institute of Technology Roorkee, Roorkee – 247667, Uttarakhand, India

[5]Indian Institute of Remote Sensing, ISRO, Dehradun - 248001, India

[6]National Atmospheric Research Laboratory, Gadanki, India.

[7]Center for Astronautical Physics and Engineering, National Central University, Taoyuan, Taiwan

[8]Department of Space Science and Engineering, National Central University, Taoyuan, Taiwan

[9]Center for Space and Remote Sensing Research, National Central University, Taoyuan, Taiwan

**Abstract**

In general, nighttime thermospheric 557.7 nm emission over mid-latitudes is predominantly masked by significantly larger mesospheric component, and hence, F-region plasma structures are rarely observed in this emission. This paper reports the first rare simultaneous detection of F-region plasma depleted structure in O($^1$D) 630.0 nm and O($^1$S) 557.7 nm airglow images from Hanle, India, a mid-latitude station (32.7°N, 78.9°E; Mlat. ~24.1°N) on a geomagnetically quiet night (Ap=3) of 26 June 2021. This indicates significant enhancement of thermospheric 557.7 nm emission. Interestingly, thermospheric 557.7 nm emission was not significant on the following geomagnetically quiet night as MSTID bands were only observed in 630.0 nm images. We show that enhanced dissociative recombination caused by descent of F-layer peak over the observation region coupled with the significant increase of the electron density at thermospheric 557.7 nm emission altitude enabled the detection of the plasma depleted structure on 26 June 2021.

**Keywords:** O($^1$D) 630.0 nm emission, O($^1$S) 557.7 nm emission, All-sky airglow imager, Dissociative recombination

**Key Points:**

1. Simultaneous observation of mid-latitude F-region plasma depleted structure in O($^1$D) 630.0 nm and O($^1$S) 557.7 nm airglow images.

2. Significantly higher electron density is observed over the region on 26 June 2021 than on the following night at airglow emission altitude.

3. Thermospheric O($^1$S) 557.7 nm emission contributed significantly due to the enhancement in the dissociative recombination reaction.

**Plain Language Summary**

The intensity of thermospheric O($^1$S) 557.7 nm emission is only ~15-20% of O($^1$D) 630.0 nm emission even though both emissions are generated via dissociative recombination reaction in the F-region. Again, the mesospheric 557.7 nm emission is produced through the Barth mechanism in the Mesosphere-Lower-Thermosphere region and this component is significantly higher than its thermospheric counterpart resulting in masking of the latter. This causes difficulties in the observation of mid-latitude F-region plasma structures simultaneously in 630.0 nm and 557.7 nm airglow images during geomagnetically quiet nights of low solar active years. To the best of our knowledge, we report for the first time, such unusual simultaneous observation of F-region plasma depleted structure from a mid-latitude station, Hanle, Ladakh, India on a geomagnetically quiet night (26 June 2021). This is a rare observation as, MSTID bands were observed only in the 630.0 images on the following geomagnetically quiet night. These observations indicate that the thermospheric 557.7 nm emission was significantly higher on 26 June than on the following night. Enhancement in the dissociative recombination due to presence of large electron density over the region as well as descent of F-layer altitude significantly increased the thermospheric 557.7 nm emission on 26 June 2021.

# 1. Introduction

The O($^1$S) 557.7 nm emission has two sources: One is in the mesosphere/lower-thermosphere (~97 ± 5 km) (Gobbi et al., 1992), and other is in the F-region/thermosphere (~250 ± 40 km) (Sobral et al., 1992). During nighttime, the mesospheric component of 557.7 nm airglow emission is generated through the Barth mechanism (Bates, 1960; Barth, 1964)

$$O(^3P) + O(^3P) + M \rightarrow O_2^* + M \quad (1)$$
$$O_2^* + O(^3P) \rightarrow O_2 + O(^1S) \quad (2)$$
$$O(^1S) \rightarrow O(^1D) + h\nu \,(557.7 \text{ nm}) \quad (3)$$

Thermospheric component of this emission as well as O($^1$D) 630.0 nm emission during nighttime are generated via dissociative recombination in the ionospheric F-region (Peterson and Van Zandt, 1969; Takahashi et al., 1990).

$$O^+ + O_2 \rightarrow O_2^+ + O \quad (4)$$
$$O_2^+ + e^- \rightarrow O^*(^1S, ^1D) + O \quad (5)$$
$$O^*(^1D) \rightarrow O(^3P) + h\nu \,(630.0 \text{ nm}) \quad (6)$$
$$O^*(^1S) \rightarrow O(^1D) + h\nu \,(557.7 \text{ nm}) \quad (7)$$

Despite having the same generation mechanism, the intensity of thermospheric 557.7 nm emission during nighttime is only ~15-20% of 630.0 nm emission (Silverman, 1970; Takahashi et al., 1990). This is, in general, true for mid-latitudes as well where the mesospheric component of 557.7 nm emission dominates its thermospheric counterpart (Shepherd et al., 1997; Swartz et al., 2000). However, at low-latitudes significant amount of 557.7 nm emission (still not dominant) originates from the thermosphere as the Appleton anomaly can generate enhanced plasma density during post-sunset hours which is sufficient to generate large thermospheric 557.7 emission through dissociative recombination (Taori et al., 2003; Rajesh et al., 2007). Although, thermospheric component of this emission is significant at low latitudes, the mesospheric component still dominates during quiet nights of solar minimum resulting in the masking of thermospheric component (Rajesh et al., 2007). At mid-latitudes where thermospheric 557.7 nm emission is not as significant as that at low-latitudes, masking effect is even more prominent (Fagundes et al., 1994; Takahashi et al., 2001). It is noteworthy that masking effect is heavily influenced by vertical advection associated with the tides and diffusion caused by the gravity waves (Shepherd et al., 1996). Due to the significant masking of thermospheric 557.7 nm emission, it is very difficult to observe any mid-latitudinal F-region/thermospheric origin plasma structures simultaneously in 557.7 nm and 630.0 nm airglow images during geomagnetically quiet nights.

Previously, a few studies reported observation of F-region plasma depleted/enhanced structures simultaneously in 630.0 nm and 557.7 nm airglow images from low-latitudes during geomagnetically active and quiet nights (Fagundes et al., 1994; Ghodpage et al., 2018; Liu et al., 2023; Sinha et al., 2001; Pimenta et al., 2004; Rajesh et al., 2007; Takahashi et al., 2001). At mid-latitudes, such type of simultaneous observation is very unusual even during geomagnetically active nights (Swartz et al., 2000). It is even rarer to observe F-region plasma structures (enhanced/depleted) in both 557.7 nm and 630.0 nm airglow images from mid-latitudes during geomagnetically quiet nights. To the best of our knowledge, such an observation has not been reported from mid-latitudes till date. During five years of airglow observations from a mid-latitude station, Hanle (32.7°N, 78.9°E; Mlat.~ 24.1°N), Ladakh, India, different types of plasma structures, viz., electrified and non-electrified medium-scale traveling ionospheric disturbances (MSTIDs), mid-latitude spread-F (MSF), and mid-latitude field-aligned plasma depletion, have been observed in 630.0 nm images (Patgiri et al., 2023a; Rathi et al., 2021, 2022, 2023; Sivakandan et al., 2020, 2021; Yadav et al., 2021a, 2021b). The present study reports a first of its kind event in which the same plasma depleted structure was observed both in 630.0 nm and 557.7 nm images on a geomagnetically quiet night (Ap=3) on 26 June of low solar active year 2021. We have analyzed multi-instrument data to uncover the reason behind this extremely rare observation.

## 2. Instruments and data analyses

In the present study, $O(^1D)$ 630.0 nm and $O(^1S)$ 557.7 nm all-sky warped airglow images as well as images without any filter captured by the imager at Hanle (32.7°N, 78.9°E; Mlat.~ 24.1°N), Ladakh, India on the night of 26 and 27 June 2021 are utilized. These warped images are obtained by removing stars and dark pixels from raw images. A detailed description of the imager including the imaging system, raw image acquisition process, geospatial calibration, methodology for image processing, and first results are discussed by Mondal et al. (2019).

Bias-corrected vertical total electron content (VTEC) from four nearby GPS receivers, KOTD (29.8°N, 78.5°E), DEHR (30.3°N, 78°E), PURO (30.9°N, 78.1°E), and DNSG (28.3°N, 83.8°E) are used to examine the deviation of VTEC associated with the plasma structures. We have also used IGS global VTEC data obtained from SPDF CDAWeb that are generated using kriging interpolation (15-minute resolution) to regional VTEC (Orús et al., 2005). The values are provided in grid of 2.5° (latitude) × 5° (longitude) and bivariate interpolation in latitude, longitude and linear time interpolation are done to calculate VTEC values at arbitrary position and time (Schaer et al. 1998). Additionally, the volume emission rate (VER) of thermospheric 557.7 nm emission is calculated by utilizing Tri-GNSS RO System electron density data from FORMOSAT-7/COSMIC-2 and neutral density data from NRL-MSISE00 model (Picone et al., 2002).

Lei et al. (2005) proposed a new solar activity proxy and a better correlation was observed between this proxy and different ionospheric parameters. Previously, many studies have used this proxy to indicate solar activity levels (Liu et al., 2007 & 2009; Li et al., 2021; Yadav et al., 2017). It is defined as

$$\Delta^{j}_{10.7P} = (\Delta^{j}_{10.7A} + \Delta^{j}_{10.7})/2.$$

Where $\Delta^{j}_{10.7A}$ is the average of previous 81 days' daily 10.7 cm wavelength solar flux values, and $\Delta^{j}_{10.7}$ represents 10.7 cm solar flux for $j^{th}$ day. In the present study, $\Delta_{10.7P}$ is used as solar activity index.

## 3. Results

### 3.1 Simultaneous observation of plasma depleted structure in 630.0 nm and 557.7 nm airglow images

Figure 1 shows a non-continuous time sequence of 630.0 nm (Figures 1a$_1$-e$_1$ & a$_2$-e$_2$), 557.7 nm (Figures 1f$_1$-j$_1$ & f$_2$-j$_2$) warped images, and images without any filter (Figures 1k$_1$-o$_1$ & k$_2$-o$_2$) captured by the Hanle imager on two successive geomagnetically quiet nights (Ap=3) of 26 June 2021 and 27 June 2021. A depleted plasma structure resembling a plume was observed in both 630.0 nm and 557.7 nm images (Figures 1a$_1$-e$_1$ & 1f$_1$-j$_1$). This plasma depleted structure was observed propagating southwestward in both sets of images with mean horizontal velocity obtained from 630.0 nm images of ~74 ± 5 ms$^{-1}$. The technique of velocity calculation of any plasma structure is described elaborately by Patgiri et al. (2023a). For checking cloud condition, images without any filter are also used (Figures 1k$_1$-o$_1$). These images clearly indicates that the plasma depleted structure observed in both filters is not contaminated by the patch of clouds visible at the southeastern edge of the imager's field of view (FOV). On the other hand, on 27 June 2021, a few northwest-southeast aligned MSTID bands with southwestward propagation (~125 ± 7 m/s) were observed in the 630.0 nm images (Figures 1a$_2$-e$_2$) while these MSTID bands were not observed in the 557.7 nm images (Figures 1f$_2$-j$_2$).

In order to check if the plasma depleted structure observed on 26 June 2021 had any low-latitude connection, we have used ionograms recorded by Gadanki digisonde (Figure s1 in supplementary file). Ionograms from this low-latitude station did not show any spread-F before, during, or after the event observed in airglow images.

**3.2 Unusually large background VTEC over the region on 26 June 2021**

Figures 2a & b show 630.0 nm airglow images on 26 June 2021 (at 15:24:45 UT) and 27 June 2021 (at 15:28:00 UT), respectively. The yellow star represents the location of the Hanle imager. The green, blue, magenta, and red curves represent the trajectories of PRN-10 at KOTD (29.8°N, 78.5°E), PRN-32 at DEHR (30.3°N, 78°E) and PURO (30.9°N, 78.1°E), and PRN-31 at DNSG (28.3°N, 83.8°E) GPS receivers for elevation angle $\geq 35°$, respectively. In both images, the locations of GPS receivers are indicated by four diamonds with the same color schemes as four trajectories. The white circle on each trajectory denotes the instantaneous location of the respective PRN on the two nights. Figures 2c-f represent temporal variations of VTEC observed by the four receivers on 26 June 2021 (solid curves) and 27 June 2021 (dashed curves). The circles on these solid and dashed curves denote the VTEC values around the airglow image times, 15:24:45 UT and 15:28:00 UT. However, there is no VTEC data for PRN 32 at PURO station after 15:12:00 UT on 27 June 2021 (Figure 2b & e). Figures 2c-e clearly show depletions in VTEC when ionospheric pierce points (IPPs) of PRNs cross the plasma structures observed in the airglow images on the two nights. It is interesting to note that the degree of depletions in VTEC observed on 26 June 2021 (~4 TECU) is much higher than the VTEC depletion observed on 27 June 2021 (~1.5 TECU). On 26 June 2021, the southern portion of the plasma depleted structure is extended to the edge of the imager's FOV (Figures 1a$_1$-e$_1$). In order to check its further southward extension, we have chosen PRN-31 at DNSG station whose trajectory lies south of the FOV (Figure 2a). The VTEC along this PRN did not show any depletion (Figure 2f) indicating the plasma depleted structure did not extend much beyond the southern edge of imager's FOV. It is interesting to note that despite being geomagnetically quiet nights, the background VTEC values observed on 26 June 2021 (~10-14.5 TECU) are significantly higher than on 27 June 2021 (~1.5-6.5 TECU). While calculating background VTEC on a particular night, we note the range of VTEC values along the four PRNs excluding the VTEC depletion region associated with the plasma depleted structure/MSTID bands. Again, the degree of depletion in VTEC for the plasma structures is calculated by subtracting the minimum VTEC from the background VTEC value (obtained just before IPPs cross the structures).

Figures 2g-j, k-n, and o-r represent global VTEC maps on 26 June 2021, 27 June 2021, and quiet time (kp<=3) average of the whole June month (24 nights excluding 26 June 2021)

over the Indian subcontinent, respectively. White stars and diamond in these maps indicate the locations of the Hanle imager and Gadanki digisonde, respectively. The purpose of taking the VTEC average of quiet nights in June 2021 is to affirm that the higher VTEC observed on 26 June 2021 is an unusual feature. It is also evident from the Figure that the overall average VTEC during 12-15 UT over the FOV of the imager was significantly higher on 26 June 2021 (~22 TECU) as compared to that on 27 June 2021 (~13 TECU) and quiet time average of whole June month (~16 ± 2 TECU).

**4. Discussion and Conclusion**

Two different types of plasma structures observed on 26 and 27 June 2021 have been critically analyzed and the structures on both nights are found to propagate southwestward having alignment along northwest-southeast (Figures 1). Despite having similar propagation direction and alignment, the degree of VTEC depletion associated with the plasma depleted structure on 26 June 2021 is significantly larger (~4 TECU) compared to that associated with the plasma depleted bands on 27 June 2021 (~1.5 TECU). It is well known that propagating MSTIDs can cause VTEC depletion of 0.5-2 TECU (Ding et al., 2011; Huang et al., 2016; Otsuka et al., 2008 & 2013; Patgiri et al., 2023a; Shiokawa et al., 2008). VTEC depletions and morphological characteristics associated with the plasma depleted bands on 27 June 2021 indicate that these bands are part of MSTID (Ding et al., 2011; Huang et al., 2016). However, the morphology and significantly large VTEC depletion associated with the plasma depleted structure on 26 June 2021 matches some of the characteristics of plasma bubbles. In order to check its low-latitude connection, we investigated the ionograms from a low-latitude station, Gadanki, which did not show any spread-F on that night (Figure s1). This indicates that the plasma depleted structure observed on this night is likely a mid-latitude plasma bubble (MPB) or mid-latitude spread-F (MSF) event. Unfortunately, we did not have any VHF radar and/or digisonde observations nearby Hanle region and, therefore, we cannot comment conclusively on whether the observed plasma depleted structure is a MPB or MSF structure.

The observation of the mid-latitude F-region plasma depleted structure both in 630.0 nm and 557.7 nm airglow images on 26 June 2021 indicates that a significant amount of 557.7 emission from the thermosphere contributed to the total 557.7 nm emission (Figures 1$a_1$-$e_1$ & $f_1$-$j_1$). On the other hand, on 27 June 2021, MSTID bands were observed only in 630.0 nm airglow images which indicates that the mesospheric component of 557.7 nm emission dominated over the thermospheric counterpart (Figures 1$a_2$-$e_2$ & $f_2$-$j_2$). One of reasons behind the enhanced thermospheric 557.7 nm emission over mid-latitude regions is the collision of oxygen atoms with super-thermal electrons (STEs) (O + e$_{STE}$ → O*($^1$S, $^1$D)) (Leonovich et al., 2012 & 2015; Tashchilin et al., 2016). Precipitation of electrons over mid-latitude regions during geomagnetic storms can increase the rate of ion production (O$_2^+$, O$^+$, etc.) and heat other thermal electrons through collisions. This, in turn, increases excitation of oxygen atoms (e.g., $^1$D, $^1$S) by thermal electrons as well as enhances the rate of dissociative recombination that ultimately increases the production 630.0 nm and thermospheric 557.7 nm emissions. The production of these emissions through the above-mentioned process is only possible during strong geomagnetic activity and at locations closer to the mid-latitude cusps. However, the year 2021 was a low solar active year ($\Delta_{10.7P}$=81.3 sfu calculated for 26 June 2021) and in particular, the two observation nights were geomagnetically quiet (Ap=3). Therefore, some other processes possibly played the role in simultaneous observation of F-region plasma depleted structure in 630.0 nm and 557.7 nm airglow images on 26 June 2021.

Thermospheric 557.7 nm emission is heavily influenced by the production of O($^1$S) state through dissociative recombination of O$_2^+$ as well as effectiveness of masking by its mesospheric counterpart (Rajesh et al., 2007; Takahashi et al., 1990). The rate of dissociative recombination reaction is dependent on the electron density (Reaction 5). On the night of 26 June 2021, we observed much higher VTEC (~14 TECU) compared to 27 June 2021 (~6 TECU) within FOV of the imager (Figures 2c-f). Additionally, in global VTEC maps, a patch of high VTEC on 26 June 2021 propagated westward through the region where we observed the plasma depletion (shown in the Movie s1 as a supplementary file) (Figures 2g-j). In addition, the electron density profiles from FORMOSAT-7/COSMIC-2 indicate the presence of significantly higher electron density on 26 June 2021 compared to 27 June 2021 (Figure 3c). These measurements on both nights were recorded closer to the airglow observation time. Tangent locations of the RO measurements with their corresponding time were also plotted over the airglow images (Figures 3a & b). Using reaction 5, we have also calculated the volume emission rate (VER) of thermospheric 557.7 nm (V$_{557.7}$) and 630.0 nm (V$_{630.0}$) emissions generated through dissociative recombination reaction (Figures 3d & e). Electron density profiles from FORMOSAT-7/COSMIC-2 and neutral density profiles from NRL-MSISE00 model are used to calculate V$_{557.7}$ and V$_{630.0}$ profiles following Sobral et al. (1992),

$$V_{557.7} = \frac{f(^1S)k_3[O_2][e]A_{557.7}}{A_{1S}\,G} \quad (8)$$

$$V_{630.0} = \frac{0.756 f(^1D)k_3[O_2][e]A_{1D}}{G(A_{1D}+k_2[N_2]+k_5[O_2]+k_6[e]+k_7[O])} \quad (9)$$

Where, $G = \frac{[e]}{[O^+]} = 1 + \left(\frac{k_3[O_2]}{k_{11}[e]+ k_4[N]}\right)\left(1 + \frac{k_4[N]}{k_{12}[e]}\right) + \frac{k_{10}[N_2]}{k_{12}[e]}$ can be deduced from charge neutrality condition (Cogger et al., 1980). Here, $[x_i]$, $A_i$, $f(^1S)/f(^1D)$, and $k_i$s denote number density of $x_i$, Einstein coefficients, quantum yield of $O(^1S)/O(^1D)$ in dissociative recombination reaction, and reaction rates (at 750°k), respectively. Magnitudes of all the coefficients used here are obtained from Sobral et al. (1992). The $V_{557.7}$ and $V_{630.0}$ are calculated using electron density profile from FORMOSAT-7/COSMIC-2. The fraction of height integrated 557.7 nm and 630.0 nm intensities $\left(\frac{\int_{h_1}^{h_2} V_{557.7}\, dh}{\int_{h_1}^{h_2} V_{630.0}\, dh} \times 100\,\%\right)$ of 557.7 nm are 14% and 11% on 26 June and 27 June 2021, respectively. This is well understood as 557.7 nm emission generated through dissociative recombination is 15-20% of 630.0 nm emission in the Thermosphere (Fagundes et al., 1994; Silverman, 1970; Takahashi et al., 1990). Interestingly, $V_{557.7}$ is significantly higher on 26 June 2021 compared to 27 June 2021 (Figure 3d). Therefore, enhanced thermospheric 557.7 nm emission on 26 June 2021 is probably due to the participation of large number of electrons in dissociative recombination process. On 26 June 2021, both electron density and VER profiles are plotted below 324 km altitude as no measurement of electron density was recorded above that altitude within the longitude and latitude range shown in Figure 3a.

It is important to note that during geomagnetically active periods, the ratio of height-integrated atomic oxygen number density to molecular nitrogen number density ($\sum[O]/[N_2]$) has been used as a diagnostic between thermospheric neutral composition and ionosphere plasma density (Burns et al., 1995a, 1995b; Cai et al., 2023; Crowley et al., 2006; Kil et al., 2011; Prolss et al., 1980; Strickland et al., 2001). Again, the storm-time modification of $\sum[O]/[N_2]$ and corresponding plasma density variations persist for more than 10 hours even during minor geomagnetic activity (Cai et al., 2020, 2021). We have checked TIMED/GUVI $\sum[O]/[N_2]$ data on 25 June 2021 (minor activity; $Kp_{max}$=3; $Dst_{min}$=-12 nT at 4 UT) and interestingly, $\sum[O]/[N_2]$ values on that day were higher compared to 26 June 2021 (~24 hour later) and 27 June 2021 (Figure s2). Contrastingly, $\sum[O]/[N_2]$ values were comparatively higher on 27 June 2021 than on 26 June 2021 over the Indian region. This neglects the role of variation of $\sum[O]/[N_2]$ caused by the geomagnetic activity on 25 June 2021 in the anomalous enhancement of electron density observed on 26 June 2021.

Besides changes in the thermospheric neutral compositions, the ionospheric altitude variation can also affect nighttime ionospheric plasma density (Farelo et al., 2002; Kil et al., 2019; Luan et al., 2013). The F-layer can be ascended/descended by equatorward/poleward meridional wind (along geomagnetic field lines) and/or eastward/westward electric field (via **E×B** drift). This causes enhancement/reduction in the F-layer electron density due to insignificant/significant chemical loss. Interestingly, in the present study, significant enhancement of electron density as well as descent of F-layer on 26 June 2021 (Figures 3c), makes the event atypical and rare. It is worth noting that the global VTEC (Movie s1) on 26 June 2021 is similar to the quiet time average except for the enhanced density beyond the equatorial ionization anomaly (EIA) crest (over the Himalayan region) around 12-15 UT. The only factor that can explain this northward transportation of plasma from the boundary of the EIA region on a geomagnetically quiet-time is probably poleward meridional wind. Generation of poleward meridional wind is possible if Equatorial Temperature and Wind Anomaly, ETWA (Raghavarao et al., 1991, 1993) is in action on 26 June 2021. Although ETWA may also be present on other days, the strength may not be enough all the time to generate stronger local poleward meridional wind as it depends on strength of EIA (Raghavarao et al., 2002; Pant & Sridharan, 2001). For investigating the role of ETWA, we have checked the strength of EIA using global VTEC near Hanle longitude (Figure s3). The EIA (northern crest) seems to be stronger on 26 June 2021 compared to 27 June 2021. Besides the peak value, TEC at Mlat ~20°-30° on 26 June 2021 is higher compared to 27 June 2021 suggesting a well-developed second EIA peak at the northern side. The second EIA peak may be a consequence of local poleward wind from the hot EIA crest due to stronger ETWA effect. This wind dragged plasma from EIA which eventually followed geomagnetic field lines to comparatively lower altitudes and overpowered the chemical loss. This may result in enhancement of electron density and descent of F-layer over the observation region. Therefore, a threshold EIA strength may exist and beyond that the ETWA effect might be strong enough to generate local poleward meridional wind resulting in the formation of second EIA peak that enhances the VER of 557.7 nm emission in the thermosphere. This study suggests that effects of ETWA in vicinity of the EIA crest need careful attention in future. However, the presence of poleward meridional wind on 26 June 2021 was not supported by either empirical model, HWM (Drop et al., 2015), or physics-based model, TIEGCM (Figures s4 & s5) outputs revealing inadequacies of these models as far as capturing low latitude ionospheric large scale plasma features are concerned.

Additionally, it is worth noting that descent of F-layer increases charge transfer reaction (Reaction 4) due to higher abundance of $O_2$ at lower altitude resulting in the enhancement of dissociative recombination of $O_2^+$ with electrons (Reaction 5) which in turn increases the thermospheric 557.7 nm emission (Bittencourt and Sahai 1979; Rajesh et al., 2007). In the present study, on 26 June 2021, we observed a significant descent of the F-layer (peak density at ~270 km) compared to 27 June 2021 (peak density at ~320 km) resulting in the enhancement of the thermospheric 557.7 nm emission (Figure 3).

Apart from enhancement in dissociative recombination, inefficient masking by the mesospheric component of 557.7 emission may play a vital role in the enhancement of the thermospheric component. However, at mid-latitude, usually mesospheric 557.7 nm emission is much higher than its thermospheric counterpart which complicates the observation of any F-region plasma structure in 557.7 nm airglow images (Fagundes et al., 1994; Rajesh et al., 2007; Takahashi et al., 2001). Even at low-latitudes, where a significant amount of 557.7 nm emission comes from thermosphere, mesospheric component of this emission still dominates the thermospheric component during low solar active year (Rajesh et al., 2007). Therefore, on a geomagnetically quiet night (26 June 2021) of low solar active year, it is quite unusual to observe any F-region plasma structure in 557.7 nm airglow images due to expected significant masking by the mesospheric component over mid-latitudes. However, there are a few other factors that can influence mesospheric masking such as eddy diffusion produced by gravity waves, and vertical advection caused by the tides (Shepherd et al., 1995 & 1996). The vertical advection associated with tides and eddy diffusion caused by gravity waves can bring oxygen rich/poor air from above/below to the altitude of maximum mesospheric emission and thereby increasing/decreasing mesospheric 557.7 nm emission (Reactions 1 and 2). Again, below a certain altitude, quenching of $O_2^*$, $O_2$, and $O(^1S)$ by neutrals dominates, resulting in the decrease of mesospheric 557.7 nm emission. It has been shown that amplitude of variation of 557.7 nm emission produced by gravity waves, in general, is not significant as compared to the variation produced by tides (Hines and Tarasick, 1987; Shepherd et al., 1996). At mid-latitude, mesospheric 557.7 nm emission shows semi-diurnal signatures that are associated with semi-diurnal tides (Petitdidier and Teitelbaum 1977, 1979). Generally, at mid-latitudes, phase of the tide is such that the emission is minimum around late evening (~20 LT). From evening onward, downward phase of the tide brings more oxygen rich air from above to the maximum airglow production layer resulting in maximum production of mesospheric 557.7 nm emission around or shortly after midnight. (Liu et al., 2008; Shepherd et al., 1996). However, evening minimum does not mean that the thermospheric component will dominate over the mesospheric one at that time. This is evident from the event observed on 27 June 2021 where the airglow observation time was around 15-16.5 UT (20.5-22 IST; IST = UT + 5.5 hr; LT = IST – 0.24 hr) in which the thermospheric signature was not observed in 557.7 nm images. On the other hand, although the observation time for the event on 26 June 2021 was around 15-16 UT, thermospheric component of 557.7 nm emission was significantly higher. As the observation periods were similar on the two consecutive nights, amplitudes of variations of mesospheric 557.7 nm emission caused by tidal oscillation on these two nights are expected to be similar. It is evident from these two events that tide is not always the dominant factor or not the only factor for large thermospheric contribution to the total 557.7 nm emission over mid-latitudes. Although this factor probably contributed to some extent to the simultaneous observation of mid-latitude F-region origin plasma depleted structure in 630.0 nm 557.7 nm airglow images on 26 June 2021, enhanced dissociative recombination reaction is possibly the most dominant factor for such rare event.

In this study, we present an event of simultaneous detection of F-region plasma depleted structure in 630.0 nm and 557.7 nm images from Hanle, Ladakh on a geomagnetically quiet night of 26 June 2021 indicating a significant enhancement in the thermospheric 557.7 nm emission. Such type of event is extremely rare as thermospheric 557.7 nm emission, in general, is insignificant as compared to its mesospheric component over mid-latitudes during geomagnetically quiet nights of low solar active years. Interestingly, on the following geomagnetically quiet night, thermospheric 557.7 nm emission was not that significant. Conclusions emerged out from various analyses indicate that descent of F-layer peak over the observational region coupled with significant increase in the electron density at the thermospheric 557.7 nm emission altitude on 26 June 2021 enhanced the dissociative recombination of $O_2^+$. This in turn magnified the production of thermospheric 557.7 nm emission which enabled to observe the plasma depletion structure in the 557.7 nm images.

**Data Availability Statement**

**Acknowledgments**

D. Patgiri acknowledges the fellowship from the Ministry of Education, Government of India for carrying out this research work. S. Sarkhel acknowledges the financial support from the Science and Engineering Research Board, Department of Science and Technology,

Government of India (CRG/2021/002052) to maintain the multi-wavelength airglow imager. The support from Indian Astronomical Observatory (operated by Indian Institute of Astrophysics, Bengaluru, India), Hanle, Ladakh, India for the day-to-day operation of the imager is duly acknowledged. R. Rathi acknowledges the fellowship from the Innovation in Science Pursuit for Inspired Research (INSPIRE) programme, Department of Science and Technology, Government of India. D. Patgiri thanks Min-Yang Chou (Research Scientist, NASA GSFC) for helping us analyzing FORMOSAT-7/COSMIC-2 data. This work is also supported by the Department of Space and the Ministry of Education, Government of India.

**Figures**

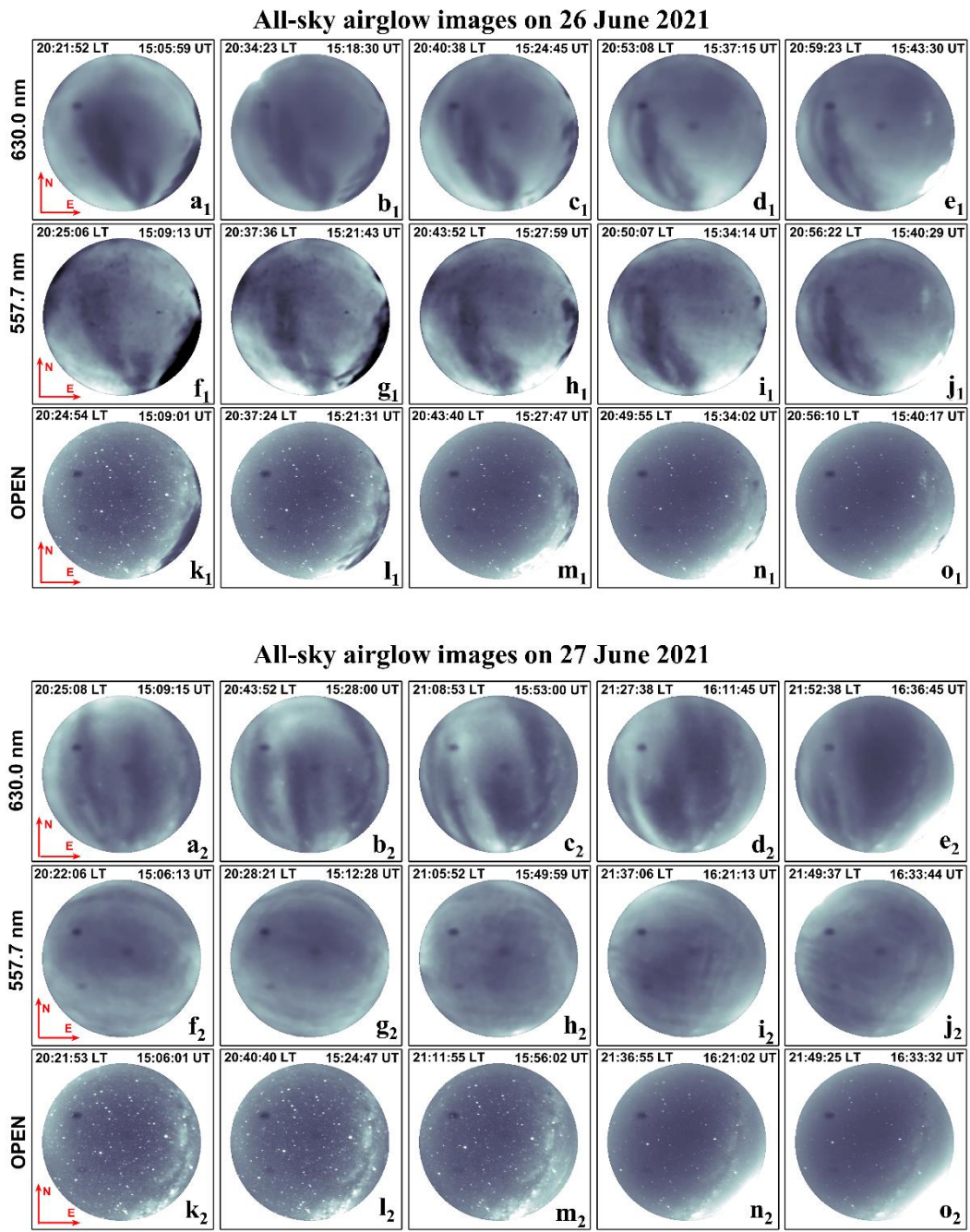

**Figure 1:** Sequences of all-sky warped airglow images on the night of 26 June 2021 ($a_1$-$o_1$) & 27 June 2021 ($a_2$-$o_2$). ($a_1$-$e_1$ & $a_2$-$e_2$) 630.0 nm images, ($f_1$-$j_1$ & $f_2$-$j_2$) 557.7 nm images, and ($k_1$-$o_1$ & $k_2$-$o_2$) images without any filter.

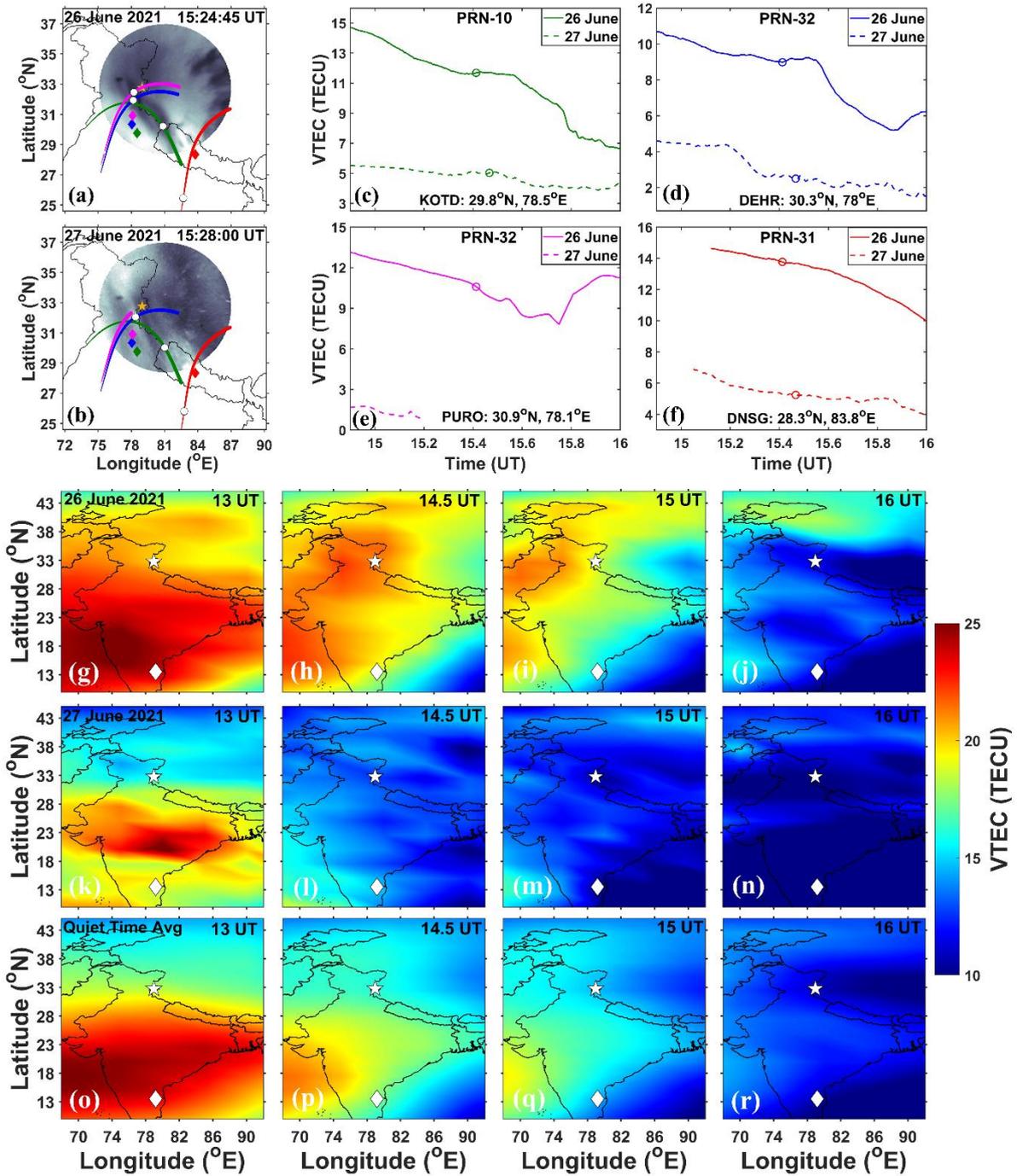

**Figure 2:** (a&b) Trajectories of PRNs observed by four GPS on 26 June 2021 (15:24:45 UT) and 27 June 2021 (15:28:00 UT). (c-f) Temporal variations of VTEC for four PRNs. (g-j, k-n, & o-r) Global TEC maps over the Indian subcontinent on 26 June 2021, 27 June 2021, & monthly quiet-time average for whole of June 2021 month (excluding 26 June 2021).

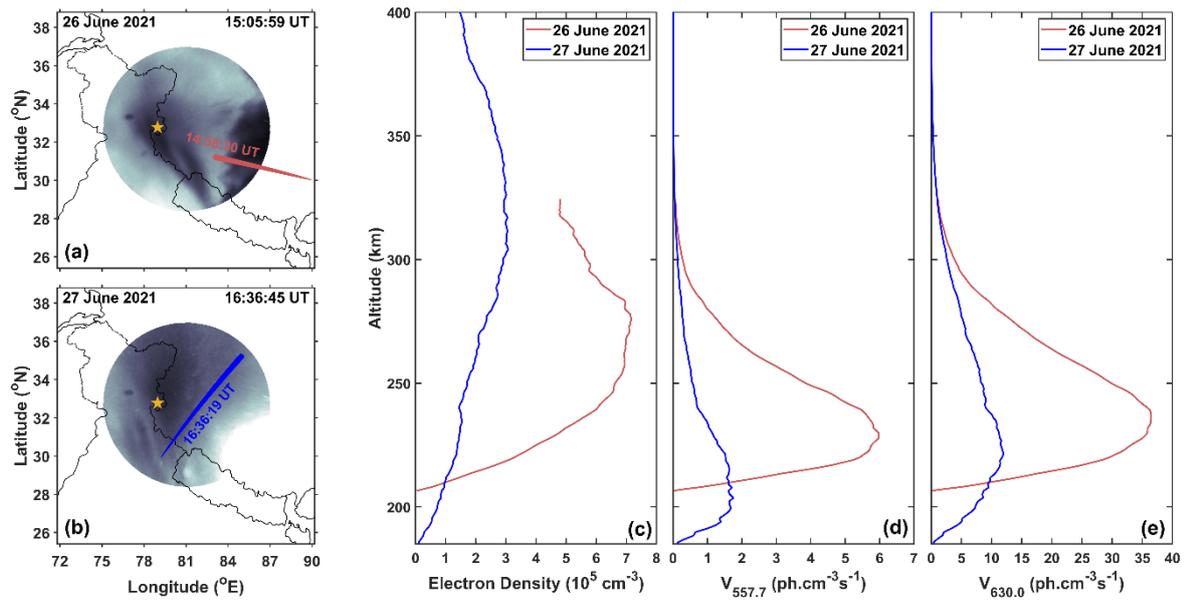

**Figure 3:** (a&b) Tangent locations of RO measurements observed by FORMOSAT-7/COSMIC-2 at 14:58:30 UT on 26 June 2021 and at 16:36:19 UT on 27 June 2021, respectively. (c,d&e) Electron density, VER profiles of thermospheric 557.7 nm, and 630.0 nm emissions for both the nights.